\documentclass[runningheads]{llncs}
\usepackage{graphicx}
\usepackage{ulem}
\usepackage{amsmath}

\newcommand{\fTrtHz}{$\mathrm{fT}/\sqrt{\mathrm{Hz}} \;$}

\newcommand{\lr}[1]{\left( {#1} \right)}
\newcommand{\xhat}{\mathbf{\hat{x}}}
\newcommand{\yhat}{\mathbf{\hat{y}}}
\newcommand{\zhat}{\mathbf{\hat{z}}}

\newcommand{\G}{\mathrm{\Gamma}}
\newcommand{\Gp}{\mathrm{\Gamma_{p}}}
\newcommand{\Gpr}{\mathrm{\Gamma_{pr}}}
\newcommand{\Gdark}{\mathrm{\Gamma_{dark}}}

\begin{document}
\title{Small animal biomagnetism applications}
\author{Kasper Jensen\inst{1} \orcidID{0000-0002-8417-4328} \and Bo Hjorth Bentzen\inst{2} \orcidID{0000-0002-0253-4992} \and Eugene S. Polzik\inst{3} \orcidID{0000-0001-9859-6591}}
\authorrunning{K. Jensen, B. H. Bentzen, and E. S. Polzik}

\institute{School of Physics and Astronomy, University of Nottingham, University Park, 
Nottingham, NG7 2RD, UK. \email{Kasper.Jensen@nottingham.ac.uk} \and
Department of Biomedical Sciences, Faculty of Health and Medical Sciences, University of Copenhagen, Blegdamsvej 3, 2200,
Copenhagen N, Denmark. \email{bobe@sund.ku.dk}
\and Niels Bohr Institute, University of Copenhagen, Blegdamsvej 17, 2100 Copenhagen, Denmark. \email{polzik@nbi.ku.dk}}
\maketitle              % typeset the header of the contribution
\begin{abstract}
The functioning of the human brain, nervous system and heart is based on the conduction of electrical signals. These electrical signals also create magnetic fields which extend outside the human body. Highly sensitive magnetometers, such as superconducting quantum interference device magnetometers or optically pumped magnetometers,  placed outside the human body can detect these biomagnetic fields and provide non-invasive measurements of e.g. brain activity, nerve impulses, and cardiac activity.
Animal models are used widely in medical research, including for disease diagnostics and for drugs testing. We review the topic of biomagnetic recordings on animal models using optically pumped magnetometers, and present our experiments on detecting nerve impulses in the frog sciatic nerve and the heart beat in an isolated guinea pig heart.
%\keywords{First keyword  \and Second keyword \and Another keyword.}
\end{abstract}

\section{Introduction} 

\subsection{High sensitivity optically pumped magnetometers for biomagnetic recordings}
An optically pumped magnetometer (OPM) is a magnetic field sensor based on optical pumping and laser-interrogation of atomic vapour such as rubidium, caesium, potassium or helium-4 \cite{Budker07,BudkerJacksonKimball2013,Fourcault2021oe}. OPMs can achieve high sensitivity in the \fTrtHz range or better \cite{Kominis2003nature,Wasilewski2010prl} and are capable of detecting small biomagnetic fields originating from the human body. OPMs can be miniaturized and, due to their operating temperature ranging from room-temperature to a few hundreds degree Celsius, it is possible to achieve a small stand-off distance of a few mm between the sensor and the skin or some other biological object.

Biomagnetic recordings have traditionally been done using cryogenically-cooled superconducting interference device (SQUID) magnetometers. For example, in magnetoencephalography (MEG), brain activity can be measured using an array of SQUID magnetometers placed in a rigid helmet. Due to the cryogenic operating temperatures of SQUIDs leading to the rigid design, MEG recordings on children have been challenging, and MEG has not been suitable for studying or examining brain activity of persons moving their head or body. 
On the other hand, OPMs can now be made compact and mounted in a flexible wearable way, which allows for MEG on children and persons performing some movement
\cite{Boto2018nature,Hill2019natcomm,Boto2019neuroimage}.

Commercial OPMs \cite{quspin} employed in neuroscience experiments utilize the spin exchange relaxation-free (SERF) magnetometer principle \cite{Allred2002prl}, where a narrow magnetometer resonance linewidth can be achieved when spin exchange collisions between the alkali atoms happen faster than their Larmor precession. SERF magnetometers are therefore operated at elevated temperatures (typically 100-200 degree Celsius) and close to zero magnetic field. Drawbacks of SERF magnetometers are their elevated operating temperature and that only signals with frequency components from DC to $\sim$100 Hz can be detected due to the zero-field condition and the narrow linewidth.
In contrast, the OPM which we have been developing is operated at room- or human body temperature and can be used to detect static, time-varying or oscillating magnetic fields with frequencies ranging from DC to MHz. The same magnetometer can therefore be used to detect slowly varying signals such as from the heartbeat (DC - 100 Hz), faster varying signals such as nerve impulses (DC - 1 kHz) as well as radio frequency magnetic fields (1-2~MHz). 

In this Chapter, we will motivate biomagnetic recordings on small animals in Sec.~\ref{sec:smallanimals}, provide some examples in Sec.~\ref{sec:examples}, and discuss ethics considerations in Sec.~\ref{sec:ethics}.
Details about our OPM and magnetometry scheme are presented in  Sec.~\ref{sec:ourOPM}.
Our work on biomagnetic applications is described in Sec.~\ref{sec:biomag}. 
We present our work on detection of nerve impulses in the frog sciatic nerve \cite{Jensen2016scirep} in Sec.~\ref{sec:nerves}. Applications to cardiology, i.e. our work on detecting the heartbeat in a guinea pig heart \cite{Jensen2018scirep} and our work towards imaging the electrical conductivity of the heart \cite{Jensen2019} are discussed in  Sec.~\ref{sec:cardiology} and \ref{sec:MIT}.

\subsection{Why biomagnetic recordings on small animals?}
\label{sec:smallanimals}
Biological processes are complex. Animals are used in biomedical research to understand the causes of diseases in both humans and animals, and to assess the safety and efficacy of novel treatment, procedures and diagnostics. This is done as it is considered wrong to deliberately expose humans to health risks, and it allows the scientists to get a better idea of the benefits and adverse events that are likely to occur once tested on humans. The use of animals is important for biomedical research. Animals are in their biology and physiology very similar to humans. Furthermore, they are susceptible to similar diseases as humans, e.g. heart diseases and cancer, and scientists can control the genetic profile of the animal, and the environment the animal is exposed to (diet, lighting, temperature etc). This is difficult to control with people. Because animals have a shorter life cycle than humans, animal models of diseases can be studied throughout the life span of the animal or across several generations, and so knowledge on how diseases progress and interact with the whole living system can be studied.      

\subsection{Examples of animal biomagnetic recordings} \label{sec:examples}
Small animal biomagnetic recordings and studies have been done using SQUID systems. For example, the magnetic field from the heart (magnetocardiography or MCG) has been studied using mice, rats and guinea pigs \cite{Brisinda2004,Miyamoto2008}.  Also the magnetic field from the brain (magnetoencephalography or MEG) has also been measured on small animals such as guinea pigs \cite{Christianson2014}.
The aforementioned advantages of a small stand-off distance and the flexible positioning of OPMs are particularly important for small animal recordings. The first detection of an animal heartbeat with an OPM was done on a mouse \cite{Lindseth2007}. Brain activity and epileptiform spikes have also been measured with an OPM in a rat \cite{Alem2014neuroscience}. Also a study of the (fetal) magnetocardiogram of chick embryos has been carried out \cite{Schellpfeffer2020}.  These measurements were performed inside a magnetically shielded environment. Unshielded measurements of the heartbeat of a cow \cite{Sutter2020elsevier} have also been carried out. Those measurements were done using two magnetometers in a gradiometer configuration together with coils for controlling the ambient magnetic field. 
Recently, OPMs based on Nitrogen-Vacancy (NV) centers in diamond \cite{Taylor2008natphys} have also been used for biomagnetic recordings. These magnetometers are solid-state, can be operated at room temperature, and are bio-compatible. Both single NV centers and ensembles of NV centers in diamond can be used for imaging magnetic fields with high spatial resolution.
Examples of biomagnetic recordings include imaging of magnetotactic bacteria with sub-micrometer resolution \cite{Sage2013nature}, detection of single giant axon action potential in marine worms and squids \cite{Barry2016pnas}, detection of action potentials in mouse muscle tissue \cite{Webb2021scirep}, and imaging of the magnetic field from the heart of a living rat \cite{Arai2021arxiv}.

\subsection{Ethics and legislation}
\label{sec:ethics}
All animal experiments are conducted in accordance with the national and internal guidelines and legislation following authorisation from the relevant authority. The legislation is established for the protection of animals used for scientific purposes. The legislation includes, among other things, rules for replacement, reduction  and refinement of procedures involving animals. This means that animal experiments are only conducted if another method for obtaining the result is not possible, and a minimum number of animals should be used. Authorisation is only given if the purposes of the project justify the use of animals and the project is designed so as to enable procedures to be carried out in the most humane and sensitive manner possible. Experiments are only to be carried out by educated and trained staff. 
In accordance with the rule of replacement of animal experiments, some biomedical questions can now be answered using computer models, in vitro work on isolated cells, cell cultures and organoids. However, nothing has yet been discovered that can function as a substitute for living organisms or complex organs. Until such discoveries, animals will still play an important role in helping scientists better understand and treat diseases.

\section{Optically pumped magnetometer} \label{sec:ourOPM}
\begin{figure}
\centering
\includegraphics[width=0.6\textwidth]{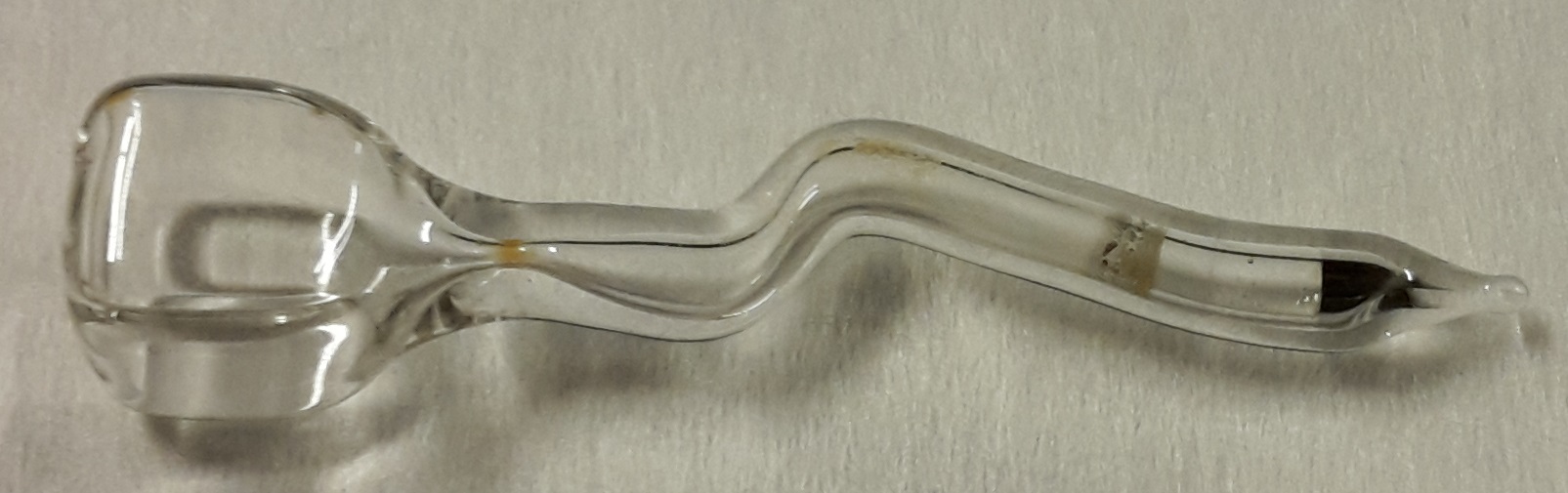}
\caption{Vapour cell containing caesium atomic vapour.} \label{fig:Cscell}
\end{figure}

We now discuss the working principle of our optically pumped magnetometer. Our OPM is based on caesium atomic vapour which is kept at room- or human body temperature in a small vapour cell.
An example of one of our hand-made vapour cells is shown in Fig.~\ref{fig:Cscell}. The vapour cell has a stem which contains a small drop of metallic caesium. Even at room-temperature, the metallic caesium partly evaporates such that the entire cell is filled with caesium vapour. The stem is connected to a 
$\left( 5~\mathrm{mm}\right)^3$ cubic volume through a small channel. Caesium atoms inside the cubic volume are optically pumped and probed using laser light.

\begin{figure}
\includegraphics[width=\textwidth]{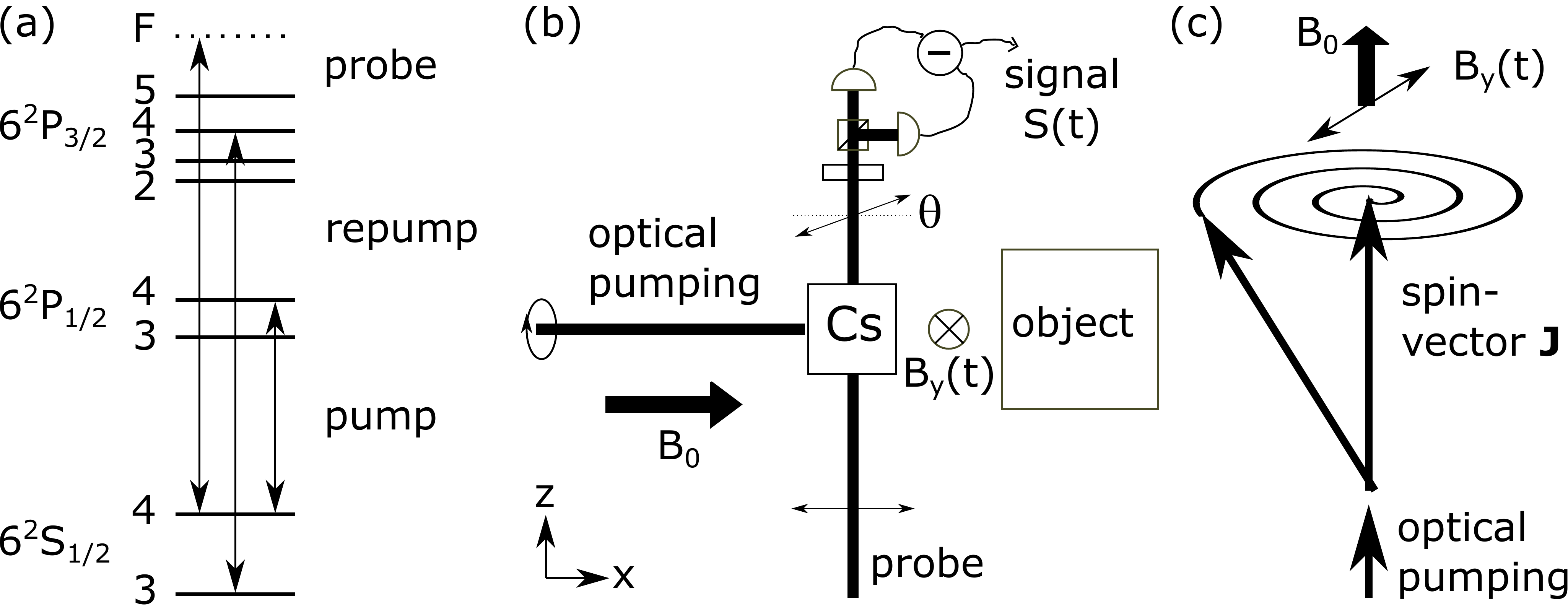}
\caption{(a) Energy level scheme of a caesium atom together with relevant optical transitions. (b) Sketch of the experimental setup. (c) Spin vector picture.} 
\label{fig:opm}
\end{figure}

The energy level scheme of a caesium atom is shown in Fig.~\ref{fig:opm}(a). The ground state is denoted $6^2\mathrm{S}_{1/2}$ and has two hyperfine levels with total angular momentum $F=3$ and $F=4$.  The first excited state $6^2\mathrm{P}_{1/2}$ has $F=3$ and $F=4$ and the second excited state $6^2\mathrm{P}_{3/2}$ has $F=2,3,4$ and 5.
We optically pump the caesium atoms with circularly polarized ``pump''  and ``repump'' light propagating along the direction of a static magnetic field $\mathbf{B}_0=B_0 \mathbf{\hat{x}}$ as shown in Fig.~\ref{fig:opm}(b).
The pump light is on resonance with the $6^2\mathrm{S}_{1/2} \rightarrow 6^2\mathrm{P}_{1/2}$ $\mathrm{D}_1$ transition in caesium and  the repump light is on resonance with the $6^2\mathrm{S}_{1/2}\rightarrow 6^2\mathrm{P}_{3/2}$ $\mathrm{D}_2$ transition in caesium as shown in Fig.~\ref{fig:opm}(a). Through optical pumping, almost all caesium atoms are prepared in the $F=4,m=4$ ground state, where $m$ is the magnetic quantum number. The quality of the optical pumping is given by the atomic polarization which can be determined by magnetic resonance measurements \cite{Julsgaard2003} and can be as high as 99.8\% \cite{Krauter2011prl}.
The ensemble of caesium atoms can be described by a spin-vector $\mathbf{J}=\left(J_x,J_y,J_z\right)$ which equals the total angular momentum of the atoms.
In general, $\mathbf{J}$ is a quantum mechanical operator satisfying the angular momentum commutation relation $\left[J_y,J_z \right]=i \hbar J_x$. 
For a fully polarised atomic ensemble the mean spin is
$\langle \mathbf{J} \rangle= \mathbf{J}_{\mathrm{max}} = 4N_A \hbar
\mathbf{\hat{x}}$. 
In our experiments where we are detecting small magnetic fields, the spin vector is only deviating slightly from the $x$-direction, such that $J_x \approx J_{\mathrm{max}}$ is approximately constant in time and the interesting dynamics is in transverse spin components $ \mathbf{J}_\perp  =\lr{J_y,J_z}$.

Caesium atoms have a magnetic moment and are affected by external magnetic fields. The combination of optical pumping along a static field $B_0 \xhat$ and the presence of a time-varying transverse magnetic field $B_y(t)\yhat$ will in general lead to rotations of the spin-vector and precession of the spin-vector at the Larmor frequency $\Omega=\gamma B_0$, as illustrated in Fig.~\ref{fig:opm}(c).
The time-evolution of the spin-vector is governed by the Bloch equation
\begin{equation}
\frac{d\mathbf{J}}{dt} = \gamma  \mathbf{J} \times \mathbf{B} + \Gp \mathbf{J}_{\mathrm{max}}-\G \mathbf{J}. \label{eq:Bloch}
\end{equation}
Here $\gamma$ is the caesium gyromagnetic ratio, $\Gp$ is the rate of optical pumping, and $\G=\Gp+\Gpr+\Gdark=1/T_2$ is the total relaxation rate of the spins which is inversely proportional to the spin-coherence time $T_2$. 
The intrinsic decay rate $\Gdark$ is due to collisions of the caesium atoms with the inner wall of the vapour cell.
At room temperature, the caesium atoms are moving with a velocity $v_{\mathrm{rms,3D}}=\sqrt{3 k_B T/m_{\mathrm{Cs}}}\approx 235$~m/s and collide with the inner cell walls around every 20~$\mu$s (for a 5~mm long vapour cell). Here $k_B$ is the Boltzmann constant, $T$ the temperature in Kelvin and $m_{\mathrm{Cs}}$ the mass of a caesium atom.
However, as the inside of our vapour cell is coated with a spin-preserving coating (alkane or alkene) \cite{Balabas2010oe}, we can achieve a long intrinsic spin-coherence time $T^{\mathrm{dark}}_2\approx 50$~ms, despite the frequent wall collisions.
Note that a minute-long spin-coherence time is possible in coated cells \cite{Balabas2010prl}. 
The decay rates due to optical pumping $\Gp$ and probing $\Gpr$ are proportional to the light power of the pump and probe light, respectively. This enables us to adjust $\G$ which is also the bandwidth of the magnetometer.

The atomic spins are probed using linearly polarized ``probe'' light which is detuned by a few GHz from the $F=4\rightarrow F=5$, $6^2\mathrm{S}_{1/2}\rightarrow 6^2\mathrm{P}_{3/2}$ $\mathrm{D}_2$ transition as shown in Fig.~\ref{fig:opm}(a).
After passing through the vapour cell, the linear polarization of the probe light is rotated by an angle $\theta$ as shown in Fig.~\ref{fig:opm}(b). The angle $\theta$ is proportional to the component of the spin-vector along the probe propagation direction, in our case $J_z(t)$. 
The angle can be measured using a balanced photodetector providing the magnetometer signal $S(t) \propto  J_z(t) $.
For simplicity, we here focus on how the mean signal changes in time $\langle S(t) \rangle \propto  \langle J_z(t) \rangle $. For a more complete quantum mechanical treatment which also describes quantum effects such as light shot noise, atomic projection noise and back-action noise from the measurements, see e.g. Ref.~\cite{Jensen2011phd}.

\begin{figure}
\includegraphics[width=\textwidth]{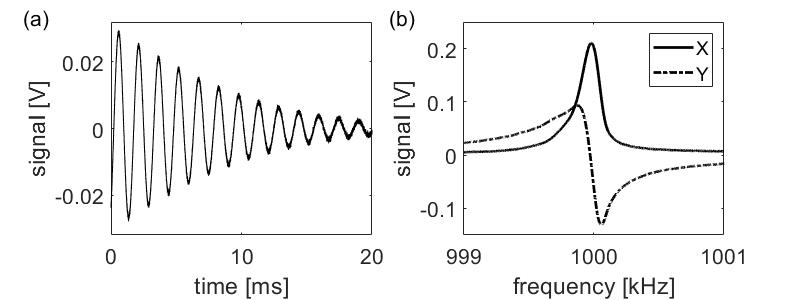}
\caption{(a) Free-induction decay. Magnetometer signal $\langle S(t) \rangle $ after a short square magnetic pulse is applied. (b) Magneto-optical resonance signal. The frequency $\omega$ of the applied oscillating magnetic field is scanned around the Larmor frequency $\Omega$ which in this case is around 1000.0~kHz. 
$X$ and $Y$ are the components of the magnetometer signal which are in-phase and out-of-phase with the applied oscillating magnetic field. }
\label{fig:FIDmors}
\end{figure}

The Bloch equation can be solved under different assumptions, which are valid for different experimental conditions. For detailed calculations see Refs.~\cite{Jensen2016scirep,Jensen2018scirep,Jensen2019}.
For example, if a short transverse magnetic pulse with duration $\tau \ll 1/\G, 1/\Omega$ is applied, then the magnetic field will rotate the spin vector creating a mean transverse spin component 
$\langle \mathbf{J}_\perp\lr{\tau} \rangle$.  
When $t\geq \tau$, the spin vector will undergo Larmor precession and decay as illustrated in Fig.~\ref{fig:opm}(c). 
The observable signal 
\begin{equation}
\langle S(t) \rangle \propto \langle J_z(t) \rangle = 
|\langle \mathbf{J}_\perp(\tau) \rangle | \cos \lr{\Omega t + \phi} e^{-\G t}
\label{eq:FID}
\end{equation}
is known as a  free-induction decay (FID). Here $\phi$ is the phase of the oscillation. 
An example FID is shown in Fig.~\ref{fig:FIDmors}(a). 
By fitting the data to Eq.~(\ref{eq:FID}), one can obtain the Larmor frequency and the decay rate which for those particular measurements were $\Omega=2\pi\cdot 650$~Hz and $\G = 1/T_2=1/\lr{8.0\;\mathrm{ms}}.$
If instead of a short square magnetic pulse, a more general time-dependent magnetic field $B_y(t)\yhat$ is applied, then the magnetometer signal will be
\begin{equation}
\langle S(t) \rangle \propto \langle J_z(t) \rangle = \gamma J_{\mathrm{max}}
\int_{t'=-\infty}^t 
\left\{ e^{-\G \lr{t-t'}} \cos \left[ \Omega \lr{t-t'} \right] \right\} B_y(t') dt'.
\end{equation}
We see that the magnetometer signal $\langle S(t) \rangle$ is a convolution of the applied magnetic field $B_y(t)$ with the  ``magnetometer response function'' $ e^{-\G t} \cos \left[ \Omega t \right] $
Once the magnetometer has been calibrated by measuring a FID and extracting $\Omega$ and $\G$, one can calculate the magnetic field $B_y(t)$ from the measured signal $S(t)$ by deconvolution. One also needs to know a scale factor which can be found by applying a magnetic field with a known amplitude. 

For the special case where the static field is close to zero $B_0 \approx 0$ and thereby the Larmor frequency $\Omega \approx 0$, the FID will be an exponential decay.
And if a more general time-dependent magnetic field $B_y(t)$ is applied, we have
\begin{equation}
\langle S(t) \rangle \propto \langle J_z(t) \rangle = 
\gamma J_{\mathrm{max}} \int_{t'=-\infty}^{t} e^{-\G \lr{t-t'}} 
B_y(t') dt',
\label{eq:lowpass}
\end{equation}
which means that the magnetometer acts as a low-pass filter.
If the transverse magnetic field is also varying slowly on a time-scale comparable to $1/\G$, then the signal is simply proportional to the magnetic field 
$ \langle S(t) \rangle \propto B_y(t)$.
As a remark, we note that in our pump-probe configuration (see Fig.~\ref{fig:opm}(b)) the magnetometer is sensitive to transverse magnetic fields. If instead the applied field is along the $z$-direction, i.e. $B_z(t)\zhat$, the response function will be  $ e^{-\G t} \sin \left[ \Omega t \right] $. In order to unambiguously extract the magnetic field from the measured signal $S(t)$, one would need to know the direction of the field, i.e. whether it is pointing along the $y$- or $z$-direction. For the special case $\Omega = 0$, the magnetometer is only sensitive to fields in the $y$-direction.

Our magnetometer can also be used to detect oscillating magnetic fields. As OPMs are based on magnetic resonance, only magnetic fields with frequency $\omega \approx \Omega$ close to the Larmor frequency will lead to a substantial signal.
In this case, one would typically process the photodetector signal with a lock-in amplifier referenced to the frequency $\omega$. The in-phase $X$ and out-of-phase $Y$ lock-in amplifier outputs then provide information about the amplitude and phase of the measured magnetic field.
In this condition, the magnetometer is often referred to as a radio-frequency magnetometer, even though the frequency in general can be anywhere in the Hz, kHz or MHz range. Figure~\ref{fig:FIDmors}(b) shows example data of the lock-in amplifier outputs while scanning the frequency $\omega$ in a 2~kHz range around 1~MHz. The in-phase and out-of-phase outputs have lorentzian and dispersive  lorentzian lineshapes, respectively, with a center frequency determined by the Larmor frequency, which in this case is around 1000.0~kHz.

\section{Biomagnetic recordings}\label{sec:biomag}

\subsection{Detection of nerve impulses } \label{sec:nerves}
Detection of action potentials and nerve impulses has scientific and clinical relevance. 
For example, measuring the conduction velocity in peripheral nerves by electrically stimulating and recording nerve activity is used for medical diagnostics. 
Magnetic field recordings could be an alternative to such electrical recordings. 
The magnetic field  generated by the axial ionic current in a signalling nerve extends outside the nerve and can therefore be detected non-invasively with a high sensitivity magnetometer.
The magnetic field from a nerve impulse was first detected by Wikswo et al.~\cite{Wikswo1980} using a SQUID magnetometer.
Here we present our work on detecting nerve impulses using a room-temperature optically pumped magnetometer \cite{Jensen2016scirep}. Although our proof-of-principle experiments were done using an isolated frog sciatic nerve, our method is compatible with \textit{in vivo} studies and could potentially have clinical applications, as also highlighted by the recent demonstration of detection of action potentials in peripheral nerves in humans using OPMs \cite{Bu2021bioarxiv}.

\begin{figure}
\centering
\includegraphics[width=0.8\textwidth]{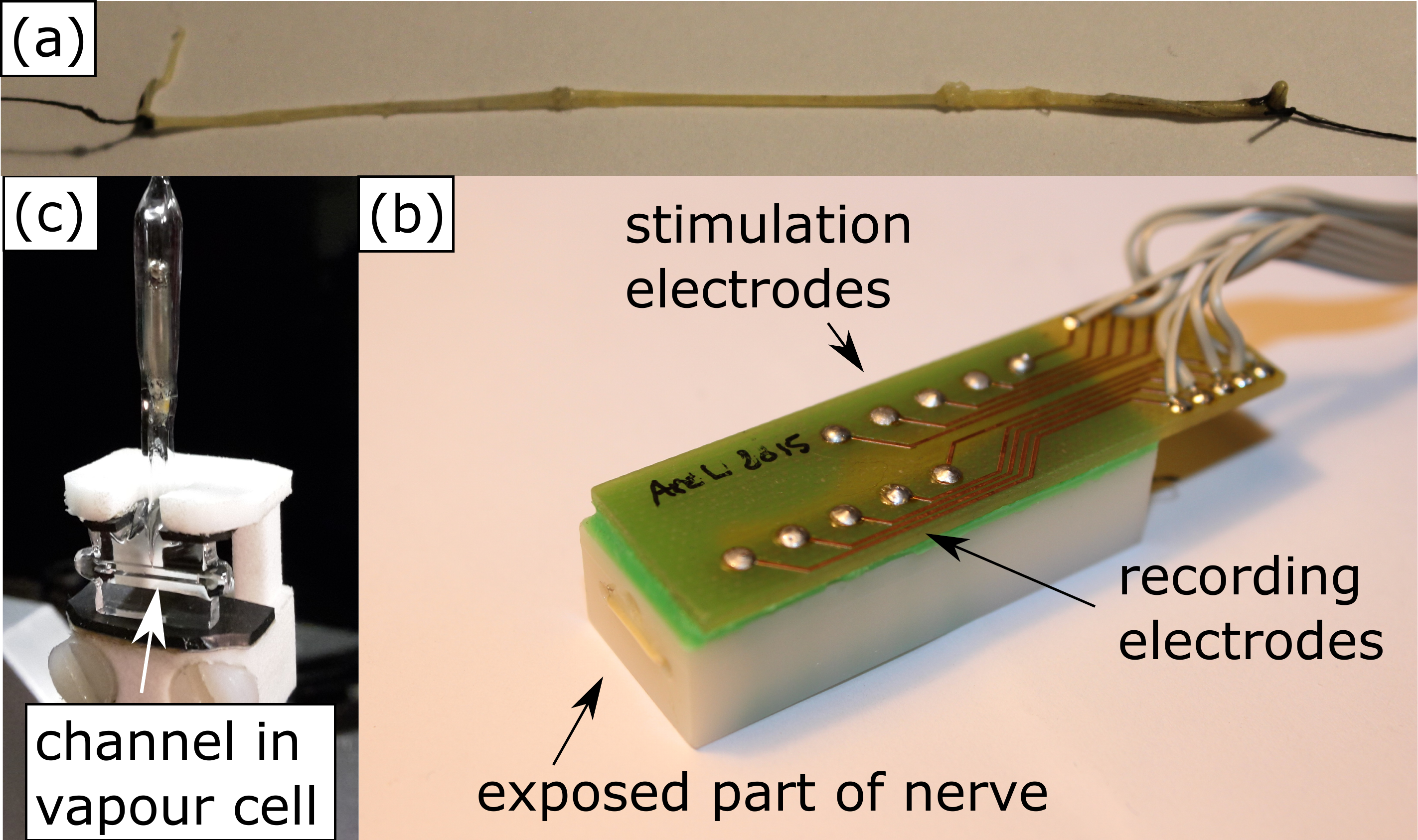}
\caption{(a) Isolated frog sciatic nerve. (b) 3D-printed plastic chamber with a hollow U-shaped channel inside. The nerve is pulled through the channel and has the middle part exposed. The chamber has  electrodes for stimulating and recording nerve impulses. The electrodes are electrically connected to the printed circuit board which has traces that take the  electrical signals to the cables at the rear. (c) Caesium vapour cell with small 1 mm x 1 mm x 7.7 mm channel inside.} 
\label{fig:nervecombined}
\end{figure}

The frog sciatic nerve is around 7-8~cm long and has 1.3~mm diameter, see Fig.~\ref{fig:nervecombined}(a). The nerve contains a few nerve bundles each with a few thousand axons inside. In our experiment, the nerve is placed in a 3D-printed plastic chamber shown in Fig~\ref{fig:nervecombined}(b). The plastic chamber contains a U-shaped hollow channel in which the nerve is pulled through. Around 1~cm of the middle part of the nerve is exposed and visible in Fig~\ref{fig:nervecombined}(b). When the nerve is kept moist inside the plastic chamber using saline solution it can stay alive for more than 5~hours. The plastic chamber has 5 electrodes on one side  and 5 electrodes on the other side. Pairs of electrodes can be used for electrical stimulation of the nerve and electrical recording of nerve impulses. The electrodes are circular with a hole in the middle (6 mm outer diameter, 1.5 mm inner diameter) and are made of gold in order to achieve a good electrical contact to the nerve. 

Magnetic field measurements are done using the caesium vapour cell shown in Fig.~\ref{fig:nervecombined}(c). The  vapour cell has a 1~mm$\times$1~mm$\times$7.7~mm channel. The magnetometer measures the average magnetic field inside that volume as only caesium atoms inside the channel are optically pumped and probed. The vapour cell is kept at room temperature and is placed close to the exposed part of the nerve with a few mm stand-off distance. As the size of the channel roughly matches the size of the exposed part of the nerve, only the magnetic field from that part of the nerve is measured.

\begin{figure}
\includegraphics[width=\textwidth]{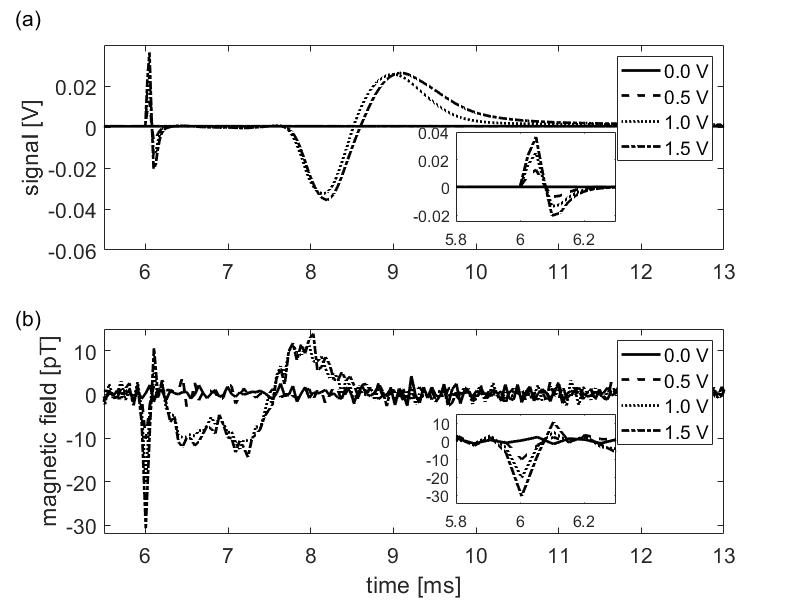}
\caption{(a) Electric and (b) magnetic recordings of nerve impulses. 
The nerve was stimulated at $t=6$~ms with a 50~$\mu$s square pulse with 0, 0.5, 1, or 1.5~V amplitude, respectively. Insets show zoom-ins of the stimulation artefacts. Data are averaged over 8000 stimulations.
} 
\label{fig:nervepulses}
\end{figure}

The magnetic field from the nerve is measured using our optically pumped magnetometer while simultaneously recording an electrical signal with a pair of recording electrodes.
At time $t=6$~ms, the nerve is stimulated with a short electrical pulse with the amplitude of either 0, 0.5, 1, or 1.5~V.
As seen in Fig.~\ref{fig:nervepulses}, nerve impulses are only propagating for the higher stimulation voltages 1 and 1.5~V. 
This is expected as the stimulation voltage should be above a certain threshold value in order to trigger a nerve impulse.
At $t=6$~ms we measure a stimulation artefact which can be differentiated from the nerve impulse by the timing and because the amplitude of the stimulation artefact is proportional to the stimulation voltage as shown in the insets in Fig.~\ref{fig:nervepulses}.
Overall, the temporal shape of the nerve impulse looks relatively similar in the magnetic and electric recordings.
We note though that the nerve impulse is detected earlier in the magnetic recording  (from $t=6-9$~ms) compared to the electric recording (from $7.5-10.5$~ms). This is because of the finite conduction velocity and because the magnetometer is placed near the middle exposed part of the nerve while the electrodes used for stimulating the nerve and recording nerve impulses are placed on opposite ends of the nerve.

\subsection{Cardiology applications} \label{sec:cardiology}

We now discuss our work on detecting the heartbeat from an isolated guinea pig heart \cite{Jensen2018scirep}. 
The guinea pig is widely used in medical research as it has biological similarities to humans. For example, the guinea pig heart has comparable size, heart rate and electrical properties as a human fetal heart of 18-22 week gestational age. Our work demonstrates that fetal-magnetocardiography (f-MCG) with OPMs \cite{Wyllie2012ol,Alem2015PhysMedBio,Eswaran2017} is promising for diagnosing heart diseases in human fetuses.

\begin{figure}
\centering
\includegraphics[width=0.8\textwidth]{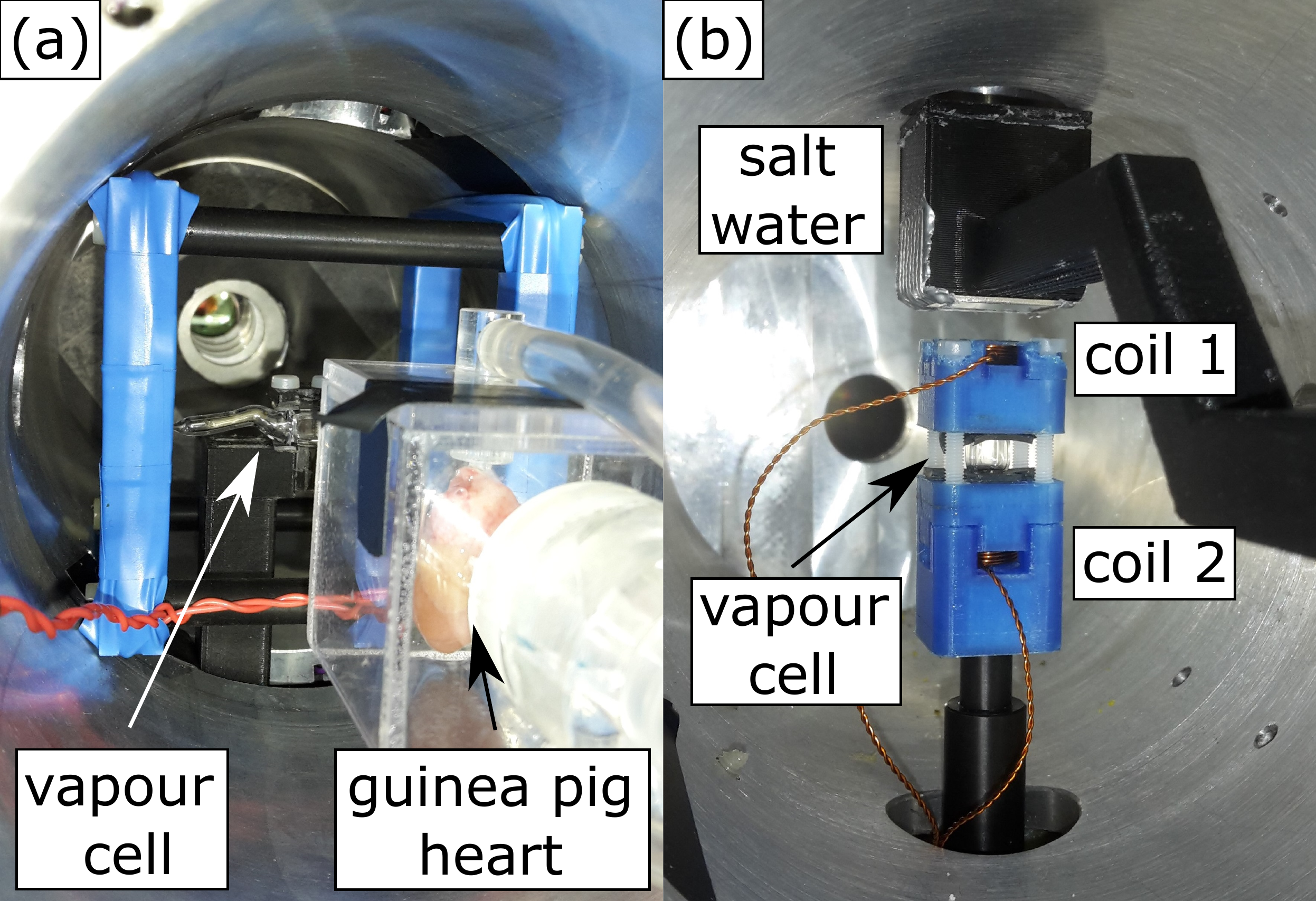}
\caption{(a) Setup for measuring the heartbeat of an isolated guinea pig heart. (b) Setup for detecting the electrical conductivity of salt-water using the magnetic induction tomography technique. } \label{fig:heart}
\end{figure}

The isolated guinea pig heart is placed in a custom-made plastic chamber and retrogradely perfused via the aorta with Krebs-Henseleit solution. This way, the heart is spontaneously beating in a regular fashion as long as the temperature, pressure and flow of solution is kept constant. The plastic chamber with the heart is placed next to the caesium vapour cell inside a magnetic shield as shown in Fig.~\ref{fig:heart}(a). The vapour cell is similar to the one shown in Fig.~\ref{fig:Cscell}, i.e. cubic with 5~mm length. 
\begin{figure}
\includegraphics[width=\textwidth]{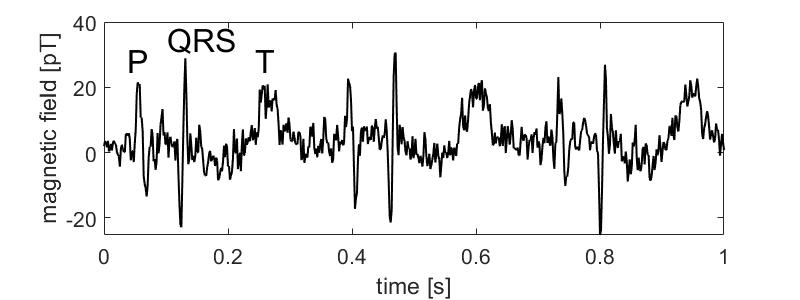}
\caption{Real-time recording of the magnetic field from an isolated guinea pig heart.} \label{fig:heartbeat}
\end{figure}
The main results of our work is that our magnetometer is capable of detecting the heartbeat from the isolated guinea pig heart in real time as shown in Fig.~\ref{fig:heartbeat}.
The heart is beating around 3 times per second and the magnetic field is around 50~pT peak-to-peak for the measurement shown Fig.~\ref{fig:heartbeat}. 
The recorded magnetic field depends on the relative position between the magnetometer and the heart. As the magnetometer (5~mm vapour cell) is smaller than the size of the heart (around 20~mm), it is possible to measure the magnetic field at different places as demonstrated in our work \cite{Jensen2018scirep}.
We achieve a good signal-to-noise ratio (SNR) of the magnetocardiogram  due to the high sensitivity of the magnetometer (120-300~\fTrtHz in the 5-1000~Hz frequency range) together with the short stand-off distance (6~mm when measured from the centre of the vapour cell to the surface of the heart). The good SNR allows us to identify the P-wave, the QRS-complex and the T-wave which are important for medical diagnostics. 
From such a real time cardiogram, one can identify if the heart is beating regularly or irregularly (e.g. if it is skipping a beat) and one can furthermore extract e.g. the RR and QT cardiac intervals. In our work 
 we also demonstrated that MCG with OPMs could be useful for diagnosing long-QT syndrome by measuring drug-induced prolongation of the QT-interval \cite{Jensen2018scirep}.

\subsection{Towards imaging the electrical conductivity of the heart} \label{sec:MIT}
We now continue with discussing our work \cite{Jensen2019} towards imaging the electrical conductivity of the heart \cite{Marmugi2016scirep} using magnetic induction tomography (MIT) \cite{Griffiths1999,Griffiths2001,Korjenevsky2000}. 
MIT is a non-invasive technique for imaging the conductivity of some object and could potentially be used for localizing areas in the heart with abnormal conductivity. This would be useful in a clinical setting for guiding ablation procedures in the heart, which are used for the treatment of e.g. atrial fibrillation. 
As a step towards MIT of the heart, we have demonstrated detection of salt-water phantoms with similar electrical conductivities as the heart \cite{Jensen2019}.  
We note that previous work towards biomedical MIT has also focused on imaging salt-water phantoms \cite{Ma2017}. MIT of the heart has not yet been demonstrated, however biomedical MIT of the spinal column has been demonstrated \cite{Feldkamp2015}.

In MIT, a coil emits a primary magnetic field $B_1(t)$ oscillating at a frequency $\omega$. 
According to Faraday's law of induction, 
the primary field induces eddy currents in any nearby conductive object, which for example could be a heart or a salt-water phantom. These eddy currents will in turn generate a secondary magnetic field $B_{\mathrm{ec}}(t)$ oscillating at the same frequency which can be detected by a nearby magnetometer. From many measurements, e.g. by scanning the relative position of the coil/magnetometer and the object, it is possible to reconstruct a 3D image of the electrical conductivity of the object. MIT of the heart is challenging for a number of reasons, including the low conductivity $\sigma \approx 1$~S/m of cardiac tissue. To address that problem, we have developed a differential technique for measuring small signals from low conductivity samples \cite{Jensen2019}.

\begin{figure}
\includegraphics[width=\textwidth]{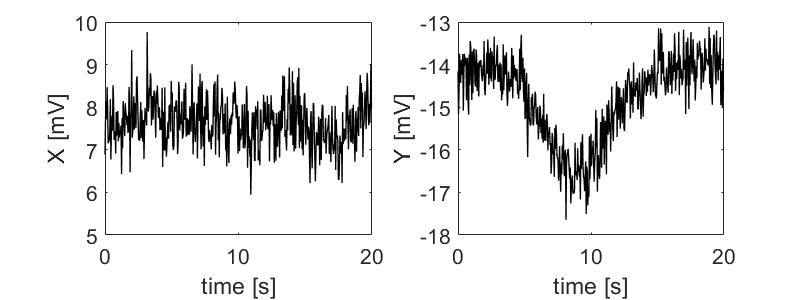}
\caption{Real time magnetometer signal when a container with salt-water with 19~S/m conductivity is scanned over coil 1. The magnetometer measures the components of the secondary field which is (a) in-phase $X$ and (b) out-of-phase $Y$ with the primary field.} \label{fig:MITresults}
\end{figure}

Our experimental setup for detecting small 8~mL salt-water phantoms using the MIT technique is shown in Fig.~\ref{fig:heart}(b). 
The conductivity of the salt-water was in the 4-24~S/m range and could be varied by changing the concentration of salt.
Coil 1 emits a primary magnetic field oscillating at 2~MHz which induces eddy currents in the salt-water phantom. 
The secondary magnetic field (averaged over the vapour cell volume) is then detected with our OPM.
A high frequency is advantageous as  $B_{\mathrm{ec}}$ is proportional to the frequency $\omega$ as long as the skin depth $\delta\lr{\omega} \approx \sqrt{2/\lr{\omega \mu_0 \sigma}}$ is much larger than the thickness of the object $t = 2$~cm. Here $\mu_0$ is the vacuum permeability. 
The secondary field is much smaller than the primary field $B_{\mathrm{ec}} \ll B_1$ due to the low conductivity. 
It is technically challenging to detect a small secondary field on top of a large primary field. Before placing the salt-water container, we therefore cancel the primary field at the magnetometer position using an auxiliary coil denoted coil 2 producing a magnetic field $B_2$. I.e., at the magnetometer position, $B_2 = -B_1$ in the absence of the object. 
We then scan the container 50~mm across coil 1 and observe a small non-zero total signal $B_{\mathrm{tot}}=B_1+B_2+B_{\mathrm{ec}}=B_{\mathrm{ec}}$. 
The differential technique allows us to detect a secondary field as small as one part in 100,000 relative to the primary field, i.e. $B_{\mathrm{ec}} = \alpha B_1$ with $\alpha = 10^{-5}$ \cite{Jensen2019}. 
Figure.~\ref{fig:MITresults} shows a real-time measurement where a container with salt-water is scanned across coil~1. The magnetometer measures the components of the secondary field which are in-phase $X$ and out-of-phase $Y$ with the primary field. When the salt-water container is directly above coil 1 (at $t=9$~ms), an out-of-phase signal is observed as expected. For this particular recording with salt-water with $\sigma=19$~S/m, the change in signal corresponds to a secondary field which is around $1/\alpha\approx5,000$ times smaller than the primary field.
Our work demonstrates that it is possible to detect low-conductivity objects with OPMs using the MIT technique. 
As MIT of the heart is a novel technique which has not yet been demonstrated, we do expect that animal studies would be appropriate. A next step would therefore be to image the electrical conductivity of a small animal heart.

\section{Conclusions}
In conclusion, we have discussed the role of small animal studies in biomagnetism and presented our own work on detecting nerve impulses and the heart beat using isolated frog nerves and guinea pig hearts. We furthermore expect that animal studies will be important for developing the magnetic induction tomography technique for imaging the electrical conductivity of biological tissue including the heart.

\section*{Acknowledgements}
This work was supported by Novo Nordisk Foundation grant NNF20OC0064182, 
by the ERC Advanced Grant Quantum-N and by the Villum Foundation

%For citations of references, we prefer the use of square brackets
%and consecutive numbers. Citations using labels or the author/year
%convention are also acceptable. The following bibliography provides
%a sample reference list with entries for journal
%articles~\cite{ref_article1}, an LNCS chapter~\cite{ref_lncs1}, a
%book~\cite{ref_book1}, proceedings without editors~\cite{ref_proc1},
%and a homepage~\cite{ref_url1}. Multiple citations are grouped
%\cite{ref_article1,ref_lncs1,ref_book1},
%\cite{ref_article1,ref_book1,ref_proc1,ref_url1}.

%\clearpage
\bibliographystyle{splncs04}
%\bibliography{references}
\bibliography{BIBOpticalMagnetometry1}

%Author, F.: Article title. Journal \textbf{2}(5), 99--110 (2016)

%Author, F., Author, S.: Title of a proceedings paper. In: Editor,
%F., Editor, S. (eds.) CONFERENCE 2016, LNCS, vol. 9999, pp. 1--13.
%Springer, Heidelberg (2016). \doi{10.10007/1234567890}

%Author, F., Author, S., Author, T.: Book title. 2nd edn. Publisher,

%Author, A.-B.: Contribution title. In: 9th International Proceedings
%on Proceedings, pp. 1--2. Publisher, Location (2010)

%LNCS Homepage, \url{http://www.springer.com/lncs}. Last accessed 4

%\end{thebibliography}
\end{document}